\documentclass[fleqn,usenatbib]{mnras}

\usepackage{newtxtext,newtxmath}
\outer\def\gtae {$\buildrel {\lower3pt\hbox{$>$}} \over 
{\lower2pt\hbox{$\sim$}} $}

\usepackage[T1]{fontenc}
\usepackage{ae,aecompl}

\usepackage{graphicx}	
\usepackage{amsmath}	
\usepackage{amssymb}	

\title[TESS observations of CD Ind]{TESS observations of the asynchronous polar CD Ind: mapping the changing accretion geometry}

\author[P. J. Hakala et al.]
{Pasi Hakala,$^{1}$\thanks{E-mail: pahakala@utu.fi (PJH)}
Gavin Ramsay,$^{2}$
Stephen B. Potter,$^{3}$
Andrew Beardmore,$^{4}$
\newauthor
David A. H. Buckley,$^{3}$
Graham Wynn$^{4}$
\\
$^{1}$Finnish Centre for Astronomy with ESO (FINCA), Quantum, University of Turku, FI-20014, Finland \\
$^{2}$Armagh Observatory \& Planetarium, College Hill, Armagh, Northern Ireland, BT61 9DG, UK\\
$^{3}$South African Astronomical Observatory, PO Box 9, Observatory, 7935, Cape Town, South Africa\\
$^{4}$Department of Physics \& Astronomy, University of Leicester, University Road, Leicester, LE1 7RH, UK.
}

\date{Accepted 2019 Apr 4; Received 2019 Apr 4; in original form 2019 Feb 28 }

\begin{document}
\label{firstpage}
\pagerange{\pageref{firstpage}--\pageref{lastpage}}
\maketitle

\begin{abstract}
We present the results of near continuous TESS optical observations of the asynchronous polar CD Ind (RX J2115-5840). The 27.9 d long light curve, with 2 min resolution, reveals remarkable changes in the magnetic accretion geometry of the system over the 7.3 d beat period. We have modelled the changes in the optical spin period pulse shape using a cyclotron emission mapping technique. The resulting cyclotron emission maps of the magnetic white dwarf reveal how the accretion geometry changes from single to two pole accretion and back over the beat cycle. Finally, we present the results from particle based numerical magnetic accretion simulations, that agree with our interpretation of the changing accretion scenario.    
\end{abstract}


\section{Introduction}

Magnetic cataclysmic variables (mCVs) are stellar binaries in which a magnetic ($B$\gtae 1 MG)
white dwarf accretes material from a late-type main sequence star through Roche Lobe overflow. The magnetic field of the white dwarf is sufficient to disrupt the formation of an accretion disc. Instead, either a truncated disc is formed or the accretion flow is channeled directly onto the magnetic pole regions of the white dwarf. Those mCVs which have $B>$10 MG are called AM Herculis systems or 'Polars', on account of their high levels of optically polarised light (see \citet{Cropper1990} and \citet{Warner1995} for reviews). 

The magnetic field strength is also high enough to lock the spin of the white dwarf with the binary orbital period. Around 100 Polars are currently known and they have orbital periods between $\sim$80 min to  8 hrs.
Of these, four (V1500 Cyg, V1432 Aql, BY Cam and CD Ind) are `asynchronous' in that the spin period is not synchronised with the orbital period. The `beat' period, $P_{b}$, is defined as $\frac{1}{P_{b}}=\frac{1}{P_{s}}-\frac{1}{P_{o}}$, where $P_{o}$ is the binary orbital period and $P_{s}$ is the spin period of the white dwarf. Over one beat cycle the accretion flow from the secondary star rotates around the magnetic field of the white dwarf, i.e. the orientation of the white dwarf magnetic field respect to the secondary star repeats at the beat period.  Depending on the orientation of the magnetic field (and its geometry) the accretion flow will follow different magnetic field lines and potentially switch from one pole to the other.

In principle, if observations can be made over a beat cycle then insight can be gained on the magnetic field geometry of the white dwarf and how the accretion stream interacts with it. In practice this is difficult for various reasons: the length of the beat cycle can be long (50 days for V1432 Aql) which will normally result in data gaps and secondly, most information is gained from polarimetry, which is often not so easily undertaken.   

CD Ind (also known as EUVE J2115--586 and RX J2115.7--5840) was discovered as a likely mCV by \citet{Vennes1996} using EUVE data and then as a X-ray source by \citet{Schwope1997}. The latter also first identified it as a possible short period asynchronous polar with the white dwarf having a magnetic field of 11$\pm$2 MG. Optical polarimetry and X-ray observations \citet{Ramsay1999,Ramsay2000} confirmed CD Ind as an asynchronous polar with a beat period of approximately a week. Using ten years of data, \citet{Myers2017} determined an orbital period of 110.812 min and a white dwarf spin period of 109.652 min, giving a beat period of 7.299 d, making it the asynchronous polar with the shortest beat period. It is therefore the most useful system to more easily study the interaction of an accretion stream with a rotating magnetic field.

\begin{figure*}
\centering
\includegraphics[width=0.8\textwidth]{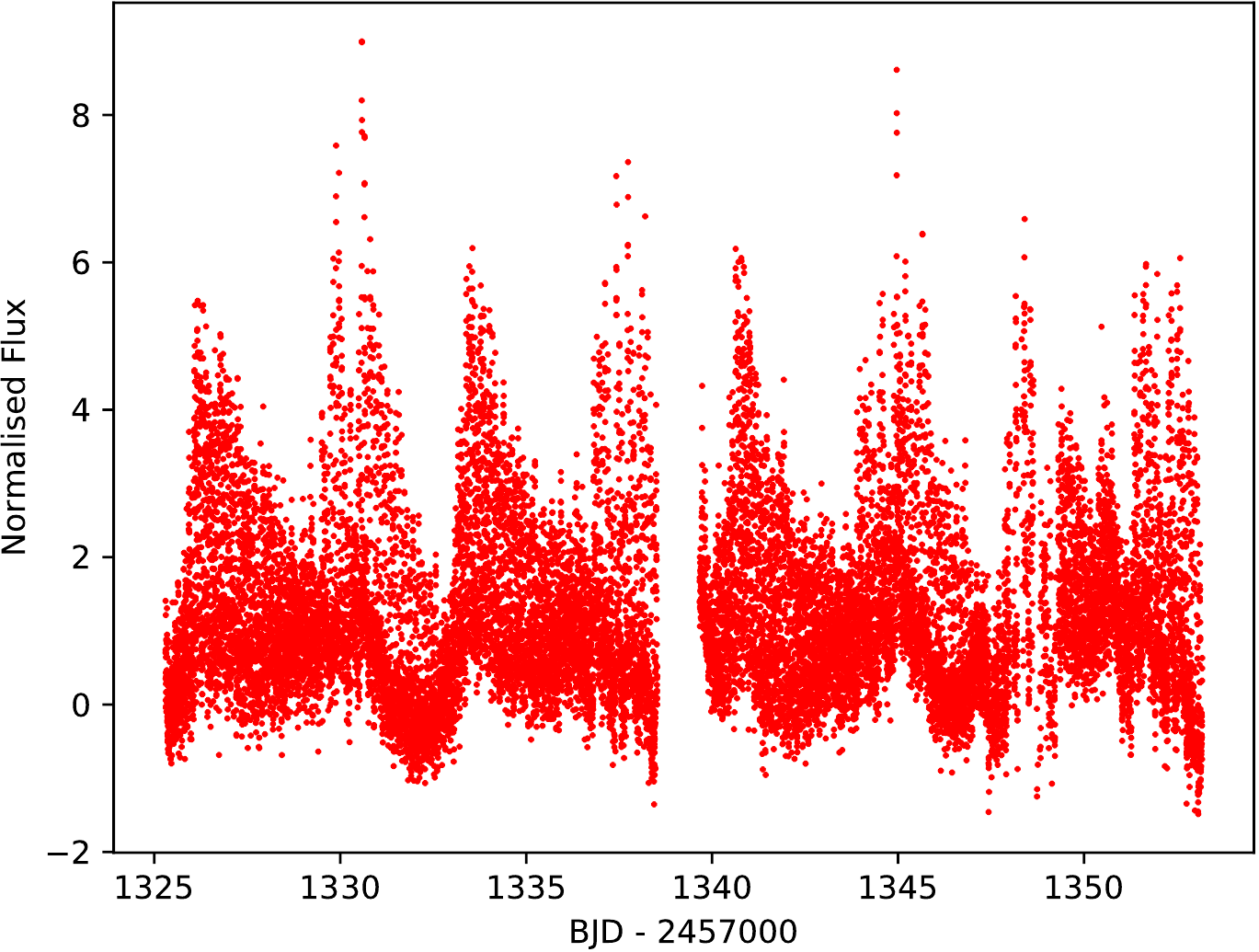}
\caption{The TESS light curve of CD Ind shown at 2 min time resolution. We have normalised the data (which has been corrected for common systematic trends) by dividing the flux of each point by the mean flux.}
\label{lightcurves}
\end{figure*}

Optical photometric observations of CD Ind have recently been obtained using the Transiting Exoplanet Survey Satellite (TESS) satellite which cover 27 days, equivalent to four beat cycles. \citet{Littlefield2019} have recently submitted a paper which uses the TESS data of CD Ind to make an in-depth study of the power spectra over the beat phase and infer the location of the accretion regions using the time of the spin and orbital maxima in the light curve. In this paper, we use the TESS data to map the accretion regions on the surface of the white dwarf using a cyclotron emission model and maximum entropy approach over the beat cycle. We also compare these maps with simulations of the accretion flow in this system.

\section{Observations}

TESS was launched on 18 April 2018 into a 13.7 d orbit. It has four small telescopes which cover an instantaneous field-of-view of 2300 square degrees (see \citep{Ricker2015} for details). Compared to Kepler it goes less deep but has a instantaneous sky coverage 20 times greater.
In contrast to Kepler, it stays on a single field for 28 days, and in contrast to K2, it will survey nearly the entire sky over the initial 2 year mission. 

The brightness of CD Ind varies considerably over the beat cycle. \citet{Myers2017} report observations of CD Ind made over ten years which show a white light magnitude in the range 15.5--18.5 mag and will therefore show a lower signal-to-noise than most of the targets which are being searched for exo-planets, the faintest being $I\sim13$ \citep{Ricker2015}. Each full-frame image is downloaded every 30 min. However, a number of targets were selected 
for higher cadence photometry, giving one photometric point every 2 min, including some which were included as a result of an open call for proposals. CD Ind was observed using TESS between 25 July and 22 August 2018 (Sector 1).

We downloaded the calibrated light curve of CD Ind from the MAST data archive\footnote{\url{https://archive.stsci.edu/tess/}}. We used the data values for {\tt PDCSAP\_FLUX} which the Simple Aperture Photometry values, {\tt SAP\_FLUX}, after it has been corrected for systematic trends common to all stars. Each photometric point is assigned a {\tt QUALITY} flag which indicates if the data may have been compromised to some degree. We removed those points which did not have {\tt QUALITY=0} and normalised it by dividing the flux of each point by the mean flux. The resulting light curve is shown in Figure \ref{lightcurves} (the gap in the middle is due to the data being downloaded to Earth at this point).

\begin{figure*}
\centering
\includegraphics[width = 0.8\textwidth]{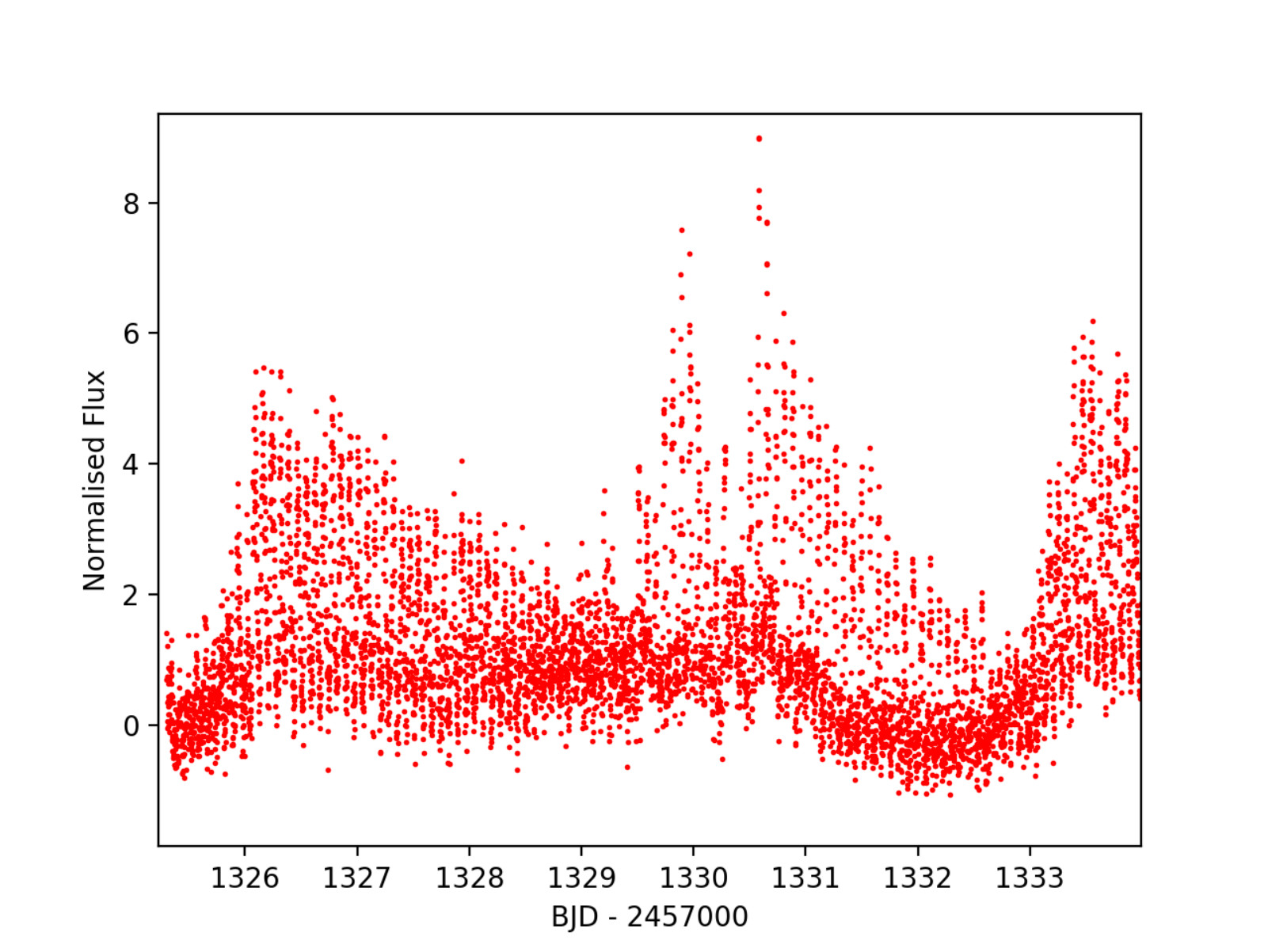}
\caption{The first eight days of the TESS light curve of CD Ind shown at 2 min time resolution. It covers one beat phase (7.3 days) and the orbital/spin period of $\sim$110 min can be seen.}
\label{lightcurvezoom}
\end{figure*}

The full light curve shows variability on many different timescales, the most obvious being the repeating features every $\sim$7.3 days. The variation in the flux corresponds to $\sim$2.5 mag, which is comparable with the variation found by \citet{Myers2017}. In Figure \ref{lightcurvezoom} we show the light curve covering the first beat cycle. On short time scales the effect of the pronounced orbital and spin period of $\sim$110 min can be seen. Our subsequent analysis uses the light curve shown in Figure \ref{lightcurves}.
 
 \begin{figure}
	\includegraphics[width=\columnwidth]{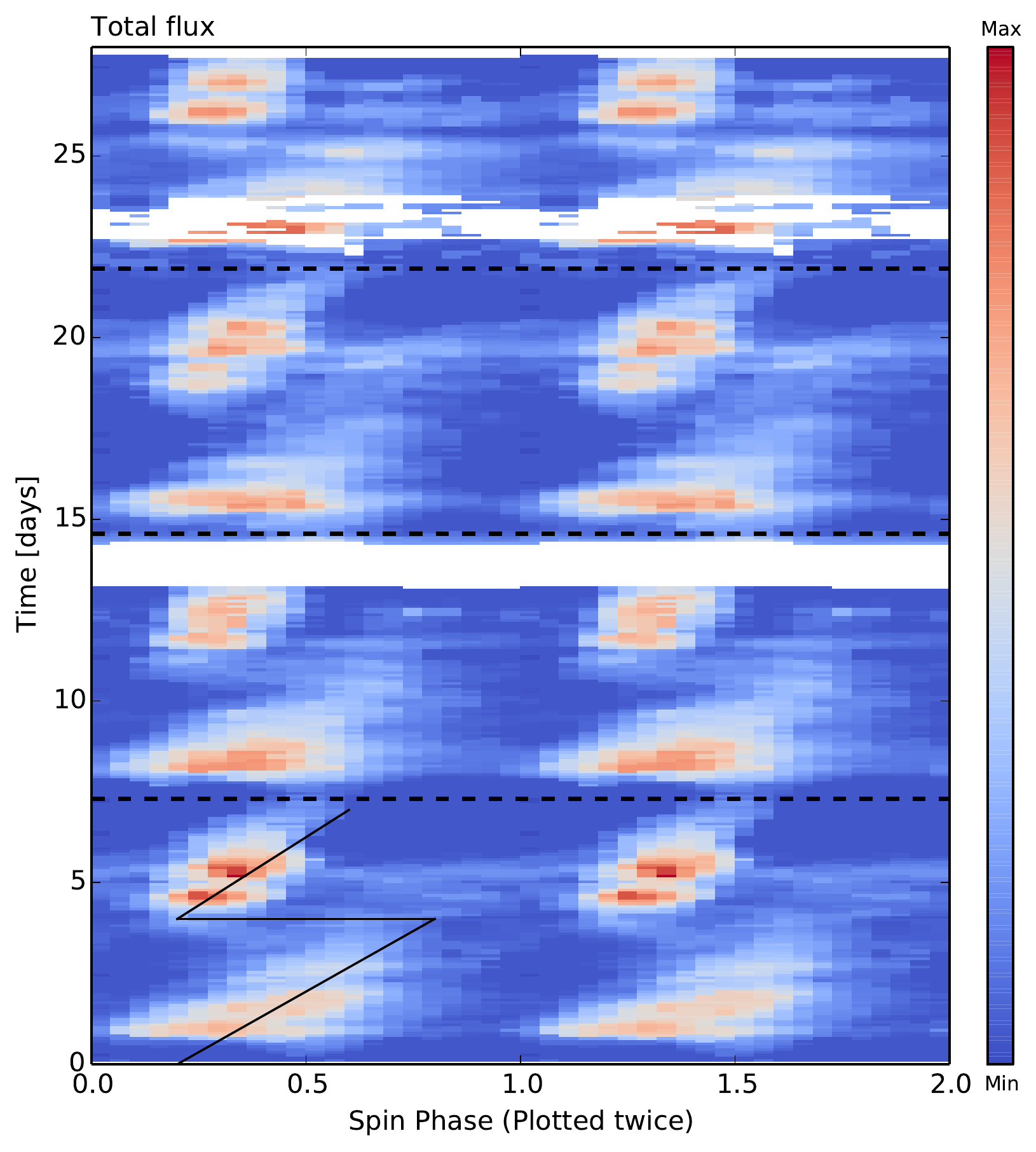}
    \caption{The evolution of the spin-folded light curve throughout the observation where time runs upwards. T0 corresponds to the time of the beginning of the TESS data set at BJD-2458325.3 which was also used as the zero point for spin and beat phasing.  The horizontal dashed lines indicate times of beat phase 0. The Z-curve shows one beat cycle of 7.3 d and how the spin pulses(s) shift over that period.}
    \label{time_spin}
\end{figure}
 
\section{Modelling of the light curves}

In our following analysis we have assumed $P_{orb}$ = 110.8124 min and $P_{spin}$ = 109.6525 min \citep{Myers2017}, which imply a beat period of 7.3 d. It should be noted, that the orbital period value (110.82005 min) quoted in \citep{Myers2017} is not compatible with the value implied by their quoted orbital frequency (110.81237 min). Moreover, their quoted spin frequency (13.1324 cyc/d, section 3.3) is not in agreement with the spin period quoted in section 3.1 (109.6564 min). For our time baseline of 27.9 d, the small difference in choice of correct spin/orbital period is not significant. 

We folded the TESS light curve of CD Ind on consecutive spin cycles and stacked them over the course of the observations (Figure ~\ref{time_spin}). The start of the observations show a broad pulse shape which advances in spin phase as a function of time. At $\sim$4 days the pulse appears to momentarily split into two local maxima after which the pulse jumps to an earlier spin phase. The broad modulation over the beat cycle is indicated by the zig-zag line in Fig.~\ref{time_spin}. This is broadly consistent with the optical intensity spin folded light curves shown in \citet{Ramsay2000}. 

\subsection{The Cyclotron mapping}

Given the clear change in the spin modulation over the beat cycle and the high quality of the data, we aim to map the cyclotron emission from the surface of the white dwarf by fitting the beat phase resolved spin light curves with a cyclotron emission model. We seek to find the most likely cyclotron emission distribution on the white dwarf surface that can fit the observed light curve at each beat phase. We have previously employed very similar mapping techniques (i.e. Stokes imaging; \citealt{PHC1998}) based on the cyclotron emission from the magnetic white dwarf surface. Of course, in  the case of TESS data, we do not have the benefit of polarisation information, which we employed in \citet{PHC1998}. This complicates the separation of cyclotron emission from opposite magnetic poles. 

Firstly, we need to identify an appropriate local cyclotron emission model to be used for the mapping. CD Ind has an mean magnetic field strength of 11$\pm$ 2 MG \citep{Schwope1997}, although there is also evidence for a non-dipole geometry, with one pole being stronger than the other \citep{Ramsay2000}. Since the wavelength of the cyclotron emission (and its harmonics) depend on the magnetic field strength, we chose to use the 10 keV, optical depth parameter $\Lambda = 10^5$ - $10^7$ and N$_{harm}$ = 8 models from \citet{WM1985} as a basis for the local cyclotron emission. The modelling was carried out with two different $\Lambda$ parameter values, as these enable us to investigate the sensitivity of our results to the chosen model. Cyclotron models with different $\Lambda$ values exhibit different strength of cyclotron beaming.  The chosen models are also consistent with the X-ray emission properties of CD Ind \citep{Ramsay2000}. These emission models produce a cyclotron hump centered in the TESS band (which covers $\sim$6000-10000\AA; \citealt{Ricker2015}). 

We are still required to make several further assumptions for the actual modelling (or cyclotron mapping). Firstly, we assume radial accretion on the white dwarf surface --- i.e. the magnetic field line orientation at the cyclotron emission 
site is parallel to the surface normal. This is strictly true only at the magnetic poles. However, our simulations (see section \S \ref{sims}) suggest that this assumption is reasonable for our purposes. It is well known, though, that in some cases non-radial accretion onto the WD seems to provide better fits to the position angle behaviour of linear polarisation from the cyclotron region(s) \citep{Cropper89}. In our case, we do not have polarisation data to verify this and as assuming some {\it ad hoc} magnetic field configuration would only introduce extra unverifiable bias in the modelling, we prefer to work with radial accretion. It should be noted that
whilst radial accretion is strictly true only at the magnetic poles, a predefined field orientation would also most certainly differ from reality in most locations over the WD surface, as indicated by an offset dipole suggested for this system by \citet{Ramsay2000}. Secondly, especially in the blue end of the optical spectrum, the accretion stream can have a large contribution to the optical spectrum. However, this should not be an important factor in the TESS band, so we can assume the cyclotron emission dominates the emission. Furthermore, the contribution by the accretion stream would not be modulated at the spin period and would appear as a diluting, constant emission component. Also, there is no evidence for emission from the secondary star in the optical spectrum \citep{Schwope1997}. Finally, in order to compute a light curve from a given cyclotron emission map, we also require the system inclination. CD Ind is not eclipsing, so the inclination must be $<75^\circ$. On the other hand, earlier studies \citep[e.g.][]{Ramsay1999,Ramsay2000} have suggested the inclination is likely to be in excess of 60$^\circ$. With these constraints in mind, we adopt a value of 70$^\circ$ for our modelling. 

\begin{figure*}
    \centering
	\includegraphics[width = 0.74\textwidth]{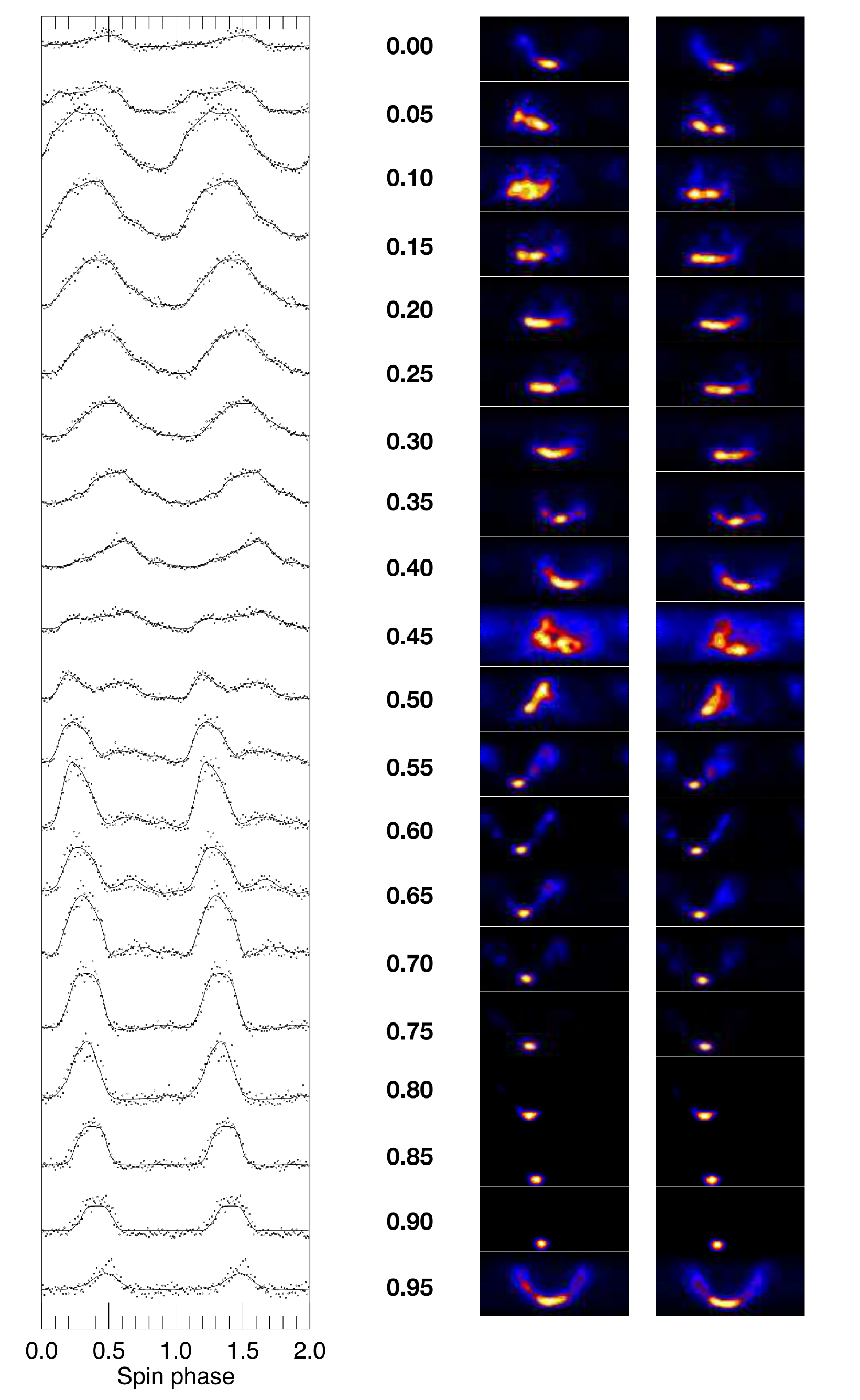}
	\includegraphics[width = 0.2445\textwidth]{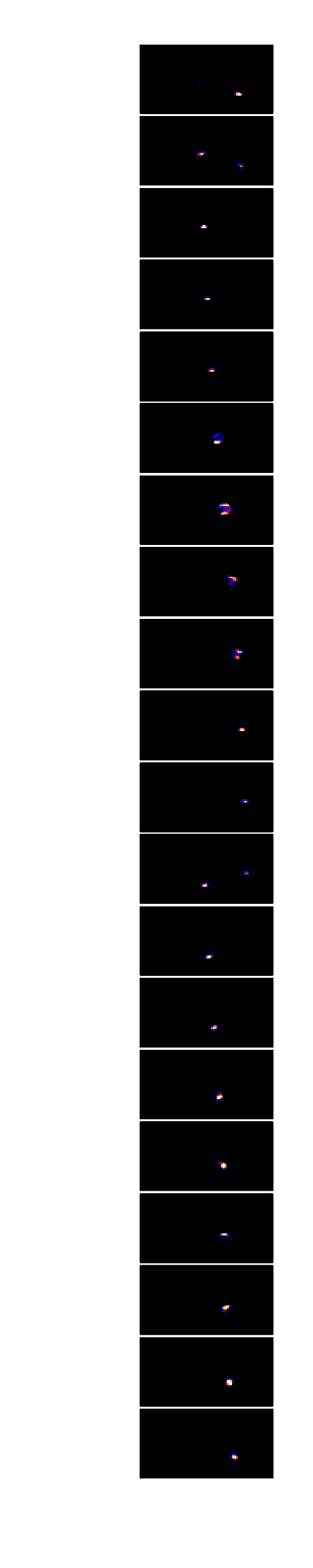}
    \caption{The spin period folded light curve as a function of the 7.3 d beat cycle, together with the best fits and corresponding cyclotron emission distributions on the white dwarf surface. The beat phase advances in steps of 0.05 by row, starting from the top. The maps corresponding to the $\Lambda=10^5$ and $\Lambda=10^7$ cyclotron model fits are shown in the second and third columns respectively. The actual model fits from the $\Lambda=10^7$ model are identical to the ones shown. The rightmost column shows the predicted accretion footprints from our numerical simulations.
    The maps cover the WD surface using the equirectangular projection (i.e longitudes 0..360 and latitudes -90..+90. equally spaced).
    Note that the beat phasing between the cyclotron maps and the accretion footprints is only approximate. Also, the longitudes of the cyclotron maps and the numerical simulation footprints are not calibrated.}
    \label{fitmaps}
\end{figure*}

\subsection{The fitting procedure}

Mapping cyclotron emission on the white dwarf surface based on light curve data is intrinsically an ill-posed problem, since the total number of data-points is far smaller than the number of free parameters (i.e. the brightness grid on the white dwarf surface). Hence, we will need to regularise our solution. The most commonly used
regularisation for similar problems is to search for maximum entropy maps that can still fit the data. We therefore seek to minimise: 

\begin{equation}
    F = \chi^2 + \lambda \sum_{i}{S_i}
\end{equation}

where $\chi^2$ is the least squares fit to the data, $\lambda$ the Lagrangian multiplier and 

\begin{equation}
S_i = p_i - q_i - p_i \cdot ln\bigg(\frac{p_i}{q_i}\bigg)
\end{equation}

is the entropy term per pixel \citep{CropperHorne1995}. $p_i$ denotes the individual pixel values and $q_i$ the default pixel value for each $p_i$, taken as the geometric mean of the four nearest pixels to $p_i$, again following \citet{CropperHorne1995}. 

In our earlier work \citep{PHC1998} we used genetic algorithms (along the lines introduced in \citet{Hakala1995} for eclipse mapping) for the Stokes imaging of the cyclotron regions. That required a separate `conventional' optimization phase to polish the final solution. Here, instead, we have adopted a global optimisation approach based on the differential evolution (DE) algorithm \citep{StornPrice1997}. DE is a population based optimisation algorithm for real-valued functions, that has been successfully used for global optimisation problems. It evolves a population of solutions based on differences within that population (with optional mutation) and tournament selection
of parent solutions that biases the next generation of solutions towards the fitter ones. The algorithm also self regulates its search space, as the differences within the population diminish during the fitting. In our implementation we have used a population of 150 cyclotron maps, that are typically evolved for 10000-30000 generations until the termination criterion is reached. This occurs when the merit function values have a standard deviation below a pre-selected limit (10$^{-5}$ of the best fit value). We refer the reader to the original work of \citet{StornPrice1997} for the detailed description of the optimisation algorithm.

\subsection{Results}

We split the light curve into twenty beat phase intervals and then phased the data in each beat phase bin on the spin period and then folded and binned the data into 100 spin phase bins. The small error on the orbital and spin periods imply that we can fold and bin the data on these periods with a very small uncertainty in the phasing ($<$0.005 in phase for the spin and even less for the orbital). The results of our cyclotron mapping are shown in Fig \ref{fitmaps} for system inclination of 70$^\circ$. They show that for the first half of the beat cycle the cyclotron emission region on the surface of the white dwarf is extended, but the second region, to which the accretion is directed during the second half of the beat cycle, is much more confined. Both emission regions are located mostly in the 'southern' hemisphere of the white dwarf. The emission maps become more complex during the pole switching phases at beat phases 0.45-0.5 and 0.95-1.0. 

\citet{Ramsay2000} used a similar approach to map the emission regions on CD Ind, but using circularly polarised optical light. These data have a more direct signature of whether the emission region is predominately in the upper or lower hemisphere. Interestingly, despite the TESS data having no polarisation information, we find that this data produces very similar results to that of \citet{Ramsay2000}. 

\begin{figure*}
\centering
\includegraphics[width = 0.79\textwidth]{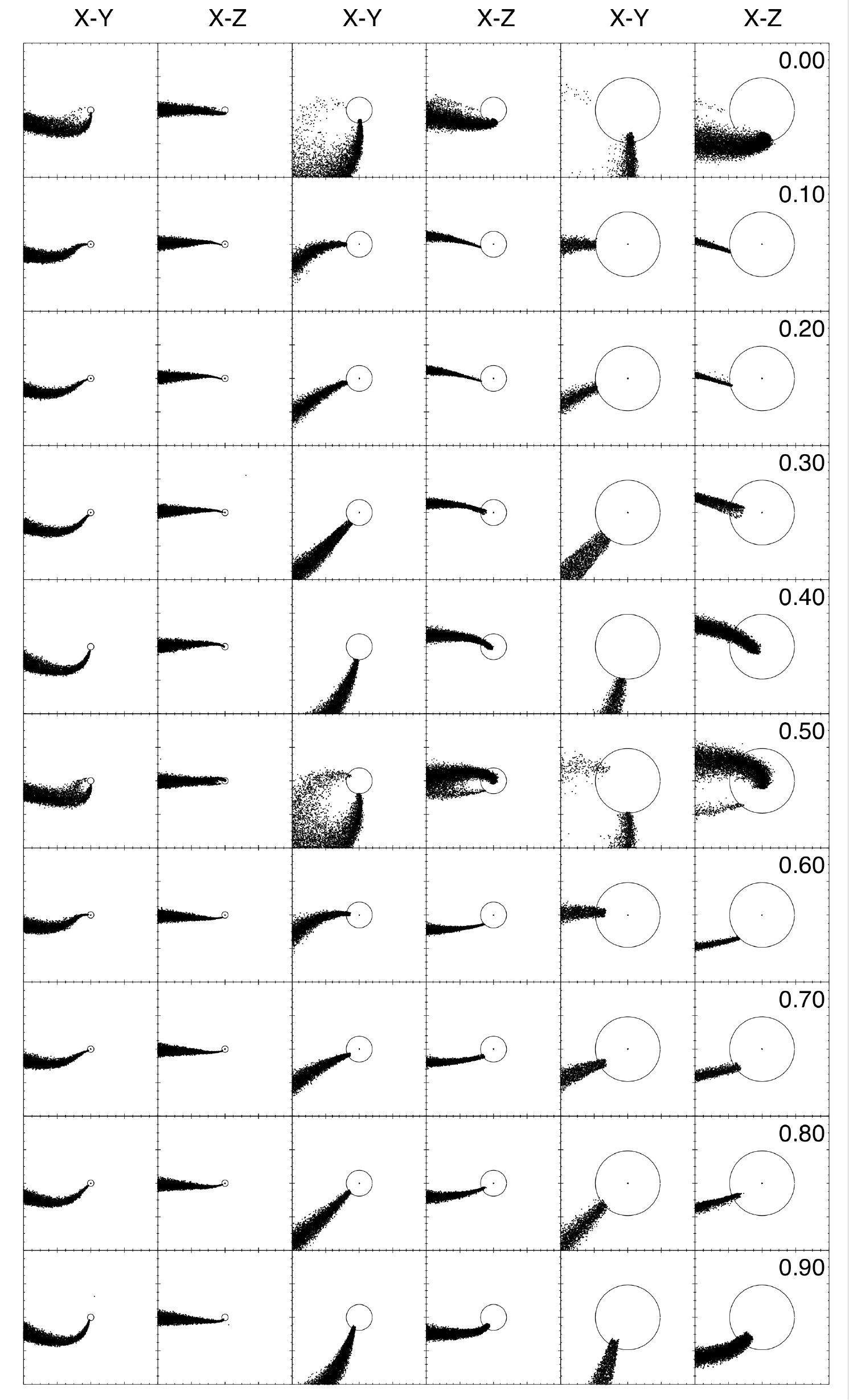}
\caption{The diamagnetic particle flow simulations in the vicinity of the WD for different beat cycle phases assuming a tilted, offset dipole geometry for the WD magnetic field. We show projections in X-Y (orbital) and X-Z plane at three different length scales (0.8$a$, 0.15$a$ and 0.08$a$ box widths, where $a$ is the binary separation). The WD radius is 0.019$a$.}
\label{accretionflows}
\end{figure*}

\section{Numerical simulations}
\label{sims}

It is fairly intuitive to think that given the small degree of asynchronism between the orbital and spin periods, the footprints of the magnetic accretion flow should not be stationary on the white dwarf surface over the beat period. Direct observational evidence for this is shown in Fig \ref{time_spin}, where the phasing of the spin pulse maxima changes over the beat period (see also \citet{Littlefield2019}). However, as the ionised accretion flow interacts with the white dwarf magnetic field, it is worth examining if such behaviour can be reproduced through simulations using our best estimate for the system parameters. With this in mind, we have carried out particle based diamagnetic 3D simulations of the mass accretion flow in the system using the {\tt HYDISC} simulation code \citep{King1993,WG1995}.

{\tt HYDISC} has been previously used for simulating magnetic accretion flows in several systems including polars, intermediate polars and the magnetic propeller system AE Aqr (see e.g. \citet{WKH1997}, \citet{Norton2008}).
At the heart of {\tt HYDISC} lies a Lagrangian, particle, or blob, based hydrodynamic code with added treatment for the magnetic interaction. This is achieved by applying a magnetic field dependent acceleration (magnetic drag) term in the equation of motion for the particles. This term, $a_{mag}$, depends on the relative shear velocity of the particles with respect to the local (rotating) magnetic field:

\begin{equation}
a_{mag} = -k({\bf r}) {\bf v_{\perp}} 
\end{equation}

where ${\bf v_{\perp}}$ is the relative shear velocity
and, in addition to that, $k({\bf r}) \propto B^2 C_A^{-1} l_b^{-1} \rho_b^{-1}$, where $C_A$ is the Alfv\'en speed in the surrounding plasma, $l_b$ is the blob length scale and $\rho_b$ is the blob density. We refer the reader to \citep{King1993,WG1995} and references therein for more detailed discussion of the simulation method.

For our purposes, we have adopted the orbital and spin periods of CD Ind as determined by \citet{Myers2017} as a basis of our simulations. Furthermore, we have assumed M$_2$=0.21$M_{\odot}$ based on the orbital period and M$_1$=0.7$M_{\odot}$ leading to the mass ratio q=0.3. The simulations were then set up using typical {\tt HYDISC} parameters for the viscosity, length scale and k-value distribution to match the magnetic field properties in Polars. A 70$^\circ$ tilted offset dipole field was fixed on the primary white dwarf and the system was evolved for several beat periods. We employed an offset of 0.5 WD radius for the dipole axis to force it ``below" the orbital plane, as suggested by the cyclotron maps.

Sample results from our simulations are shown in Figure \ref{accretionflows}. The beat phases of the chosen samples are 
shown. These correspond approximately to the phasing shown in 
Fig \ref{fitmaps}. The six different columns show the results in X-Y (orbital) and X-Z planes at three different length scales, as indicated by the size of the WD. It is clearly evident that the accretion stream 
 flips between the two magnetic poles, and during pole switching, a short duration ($\sim$ 0.05 in beat phase)  double  pole accretion phase also takes place. The radially offset dipole directs the accretion flow to two different, predominantly `southern', regions.        

\begin{figure}
    \centering
	\includegraphics[width = 0.45\textwidth]{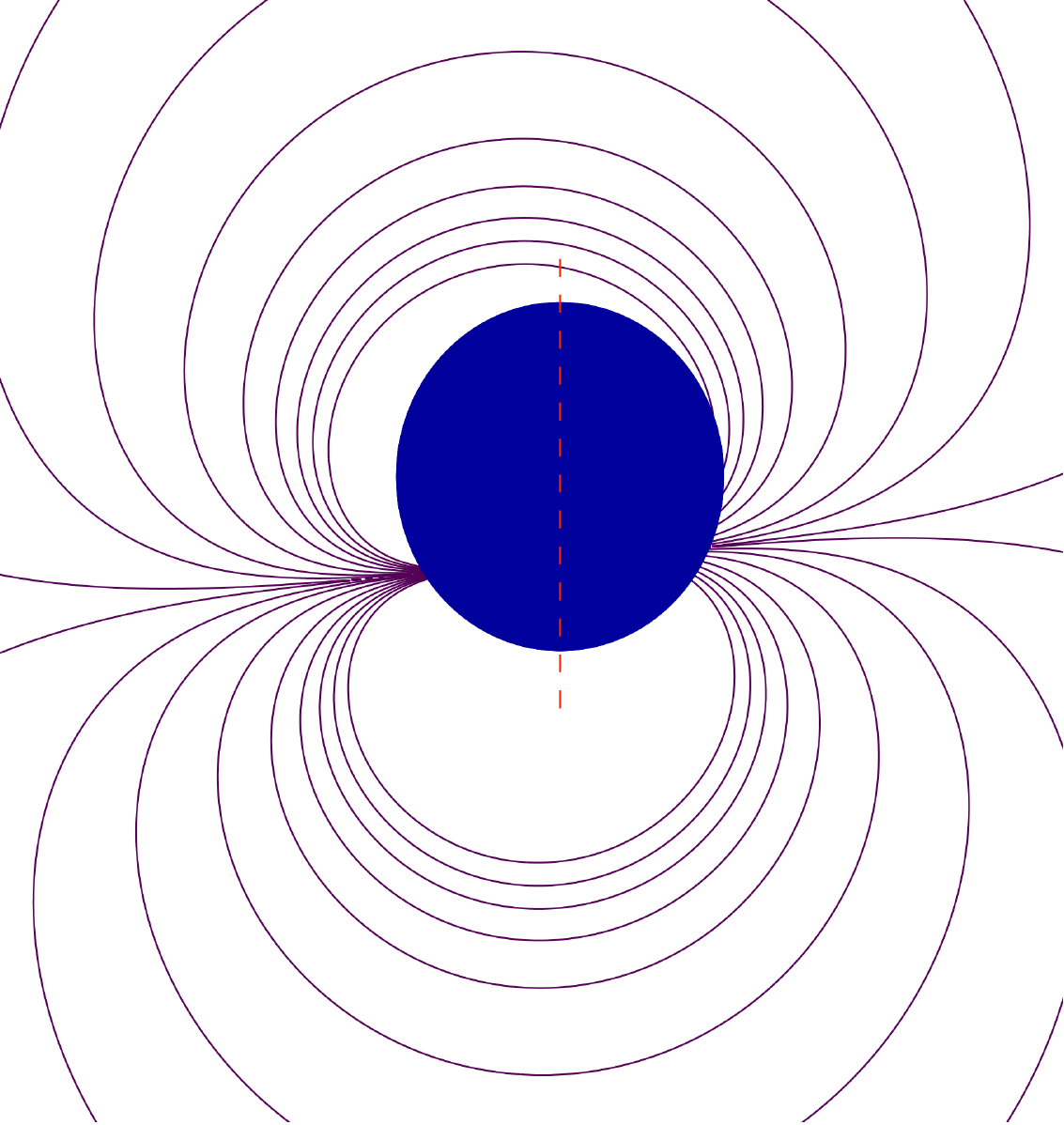}
    \caption{A schematic view of the proposed magnetic field configuration based on the resulting cyclotron maps. The vertical hashed line is the spin axis and the WD is viewed in the orbital plane.}
    \label{wdfield}
\end{figure}

\section{Discussion}

\citet{Ramsay1999} analysed circular polarimetry data taken over two weeks. They found that over the beat cycle the sign of the circular polarisation reversed, indicating that the accretion stream changed from accreting onto one and then the other magnetic pole. They considered two scenario's, one in which the accretion flow from the secondary star is directed onto one or other footprint of the same set of field lines over the whole beat cycle. In the second scenario the flow is directed on to roughly diametrically opposite field lines at different beat phases. Based on observations of the spin phase of the positive and negative polarisation peaks, \citet{Ramsay1999} concluded that the first  scenario was much more likely (see Fig 5 of \citealt{Ramsay1999}). Additional polarimetry and X-ray data allowed \citet{Ramsay2000} to conclude that white dwarf had a non-dipole field with the upper pole having a significantly higher magnetic field strength than the lower pole. 

Our modelling of cyclotron emission over the WD surface does not pick any clearly significant emission located in the `northern' hemisphere. Instead, during the first half of the beat cycle, we detect an extended emission arc/curtain in the `southern' hemisphere, which is replaced by emission from a more compact (and more `southern') region during the second half of the beat period. There is some indication for a latitudinally extended emission during the beat phases 0.95-1.05 and 0.40-0.50, when the pole switching takes place (Fig \ref{fitmaps}). Our trials with inversion of synthetic maps that included both `northern' and `southern' emission regions clearly showed that should a `northern' region exist, it would be picked up by the modelling. This is, however, not observed in our real data maps (Fig \ref{fitmaps}). 

Since both accreting regions appear to be in the `southern' hemisphere, but at somewhat different latitudes, we consider a model, where the (presumably) dipole field of the WD is tilted 70$^\circ$ from the spin axis and also offset by 0.5 R$_{WD}$ from the center of the WD (Fig. \ref{wdfield}). This would produce two `southern' emission regions that could account for the observed cyclotron maps. Indeed, our numerical simulations with such a field geometry manage to produce the expected locations for the accretion regions. However, we are not able to reproduce the pronounced longitudinal extent difference in emission from the two poles, which is suggested by the cyclotron maps (Fig. \ref{fitmaps}.). The difference of spin pulse profiles from the two poles is real though, as it can be clearly seen in Fig. \ref{time_spin} and Fig. \ref{fitmaps} that the spin period folded light curves in the beat phase ranges 0.1-0.4 and 0.6-0.9 are very different. Explaining this might require a dipole field that is also offset along its magnetic axis, or a more complicated field geometry.

It is also apparent that the accretion maps based on the numerical simulations (Fig. 5) produce much more confined accretion footprints on the WD surface than our cyclotron modelling (Fig. 4) implies.
 There are at least two possible reasons for this. Firstly, the maximum entropy regularisation applied to the inversion of the light curves, together with the chosen pixel grid resolution, smooths out the resulting maps to some extent. Secondly, the accretion footprints based on the numerical simulations depend to some extent on the chosen
strength and distribution of the particle-by-particle magnetic drag force coefficient, $k$. Wider distribution of $k$-coefficients would lead to more extended accretion arcs on the WD surface. Both the cyclotron maps and the accretion footprints as a function of the beat phase are shown in Fig. \ref{fitmaps}.

\section{Conclusions}

We have employed a near continuous (over 27.9 d) TESS high cadence light curve of CD Ind in order to study the changes in the magnetic accretion process over several beat cycles of this asynchronous polar. Our cyclotron emission maps, based on evolutionary maximum entropy inversion show that accretion onto the magnetic white dwarf primary flips between two magnetic poles. Furthermore, the accretion footprints seem to closely follow the locations of magnetic poles over the beat cycle, before the stream trajectory switches to the other pole. Double pole accretion occurs for a maximum of 0.05 in beat phase during each of the pole transitions.
This is confirmed by both the numerical simulations and the cyclotron mapping. We conclude that either the system contains a tilted offset dipole, the axis of which lies entirely below the orbital plane (Fig \ref{wdfield}.) or alternatively a higher order magnetic field configuration is required to account for the accretion geometry and its changes over the beat period.   

\section*{Acknowledgements}

This paper includes data collected by the TESS mission, which are publicly available from the Mikulski Archive for Space Telescopes (MAST). Funding for the TESS mission is provided by NASA's Science Mission directorate. Armagh Observatory and Planetarium is core funded by the Northern Ireland Executive. SBP and DAHB acknowledge support of the National Research Foundation of South Africa. APB acknowledges support from the UK Space Agency. Since the submission of this paper a preprint by Littlefield et al. (2019) was submitted to the http://arxiv.org, where they analyse the same dataset and suggest a different identification for the spin and orbital periods. We thank Colin Littlefield for communicating their results in advance.

\bsp	
\label{lastpage}
\end{document}